\documentclass[a4paper,11pt]{article}
\pdfoutput=1
\usepackage{jheppub}
\usepackage{textcase}
\usepackage{graphicx,epsfig,psfrag,amssymb,hyperref}
\usepackage{multirow}
\usepackage{color,graphicx,epsfig,psfrag,amsmath,empheq}
\usepackage[dvipsnames]{xcolor}
\usepackage{bm}
\usepackage{mathrsfs,amsfonts,soul,color}
\usepackage[caption=false]{subfig}
\usepackage{hepunits}
\usepackage{enumitem}
\usepackage{placeins}
\usepackage{textcomp}
\usepackage{gensymb}

\let\pt=\pT

\begin{document}

\title{Beyond 4D Tracking: \\ Using Cluster Shapes for Track Seeding}

\author{Patrick J.~Fox$^a$,}
\author{Shangqing Huang$^{b,c}$,}
\author{Joshua Isaacson$^a$,}
\author{Xiangyang Ju$^c$, \\ and}
\author{Benjamin Nachman$^{c,d}$}

\affiliation{
\begin{scriptsize}
\phantom{ }\hspace{-0.12in}$^a$Theoretical Physics Department, Fermi National Accelerator Laboratory, P.O. Box 500, Batavia, IL 60510, USA \\
\phantom{ }\hspace{-0.12in}$^b$Department of Physics, University of California Berkeley, Berkeley, CA 94720, USA \\
\phantom{ }\hspace{-0.12in}$^c$Physics Division, Lawrence Berkeley National Laboratory, Berkeley, CA 94720, USA \\
\phantom{ }\hspace{-0.12in}$^d$Berkeley Institute for Data Science, University of California, Berkeley, CA 94720, USA \\
\end{scriptsize}
}

\emailAdd{pjfox@fnal.gov}
\emailAdd{shangqing@berkeley.edu}
\emailAdd{isaacson@fnal.gov}
\emailAdd{xju@lbl.gov}
\emailAdd{bpnachman@lbl.gov}

\preprint{FERMILAB-PUB-20-650-T}

\abstract{
Tracking is one of the most time consuming aspects of event reconstruction at the Large Hadron Collider (LHC) and its high-luminosity upgrade (HL-LHC).  Innovative detector technologies extend tracking to four-dimensions by including timing in the pattern recognition and parameter estimation.  However, present and future hardware already have additional information that is largely unused by existing track seeding algorithms.  The shape of pixel-clusters provides an additional dimension for track seeding that can significantly reduce the combinatorial challenge of track finding.  We use neural networks to show that cluster shapes can reduce significantly the rate of fake combinatorical backgrounds while preserving a high efficiency.  We demonstrate this using the information in  cluster singlets, doublets and triplets.  Numerical results are presented with simulations from the TrackML challenge.
}

\maketitle

\section{Introduction}
\label{sec:intro}

Analyzing data from the Large Hadron Collider (LHC) present a \textit{hyper} challenge.  A given collision event may result in hundreds of outgoing particles, each with many features (momentum, electric charge, etc.). This \textit{hyper}variate phase space is then observed by complex multi-channel detectors that are essentially hyperspectral cameras.  The LHC detectors have millions of readout channels and dimensionality reduction is essential for data analysis.  One natural and nearly lossless reduction is the reconstruction of charged particle trajectories (`tracks').  The innermost layers of the detectors at the LHC are constructed to register the passage of charged particles without significantly altering the particle energy or direction.  In the ATLAS and CMS detectors, this is achieved using silicon sensors that are finely segmented in one or two directions and are called strips and pixels, respectively.  We will focus on pixels, although our methodology applies more generally.

Typically, the first step in a tracking algorithm is the construction of seeds, which are sets of three or more hit pixel clusters that can be used to fit charged-particle trajectories (see e.g. Ref.~\cite{Aaboud:2017all,Chatrchyan:2014fea}).  If there are $\mathcal{O}(10)$ detector layers, a given event with $\mathcal{O}(10^3)$ particles would have $\mathcal{O}(10^4)$ pixel clusters and therefore about $\mathcal{O}(10^{10})$ possible seeds.  Since the number of seeds per real particle is $\mathcal{O}(1)$, the initial purity of seeds can be as low as $\mathcal{O}(10^{-10})$.  This is a significant challenge for pattern recognition at the LHC and its high-luminosity upgrade, the HL-LHC, where the number of particles per event will grow even higher.

The geometric arrangement of seeds in space can be used to eliminate those that are unlikely the result of a real particle.  However, given the significant challenge presented by the large number of seeds, it is important to examine other non-geometric solutions.  One solution is to augment tracking detectors with precise timing information to  reject combinations of clusters from different collisions~\cite{Sadrozinski:2017qpv,CERN-LHCC-2017-027,CERN-LHCC-2018-023}.  This is especially effective at high luminosity where a given proton-proton bunch crossing may result in $\mathcal{O}(100)$ individual proton-proton collisions with a non-negligible spread in time.  While timing is a promising avenue, the existing detectors are not fast enough for this information to be useful.  Other hardware modifications have been studied, such as linking two nearby spatial hits~\cite{CERN-LHCC-2017-009}, but this is also not available in existing detectors and is currently not possible for the innermost detector layers due to bandwidth limitations.

\begin{figure}[t]
    \centering
    \includegraphics[width=0.95\textwidth]{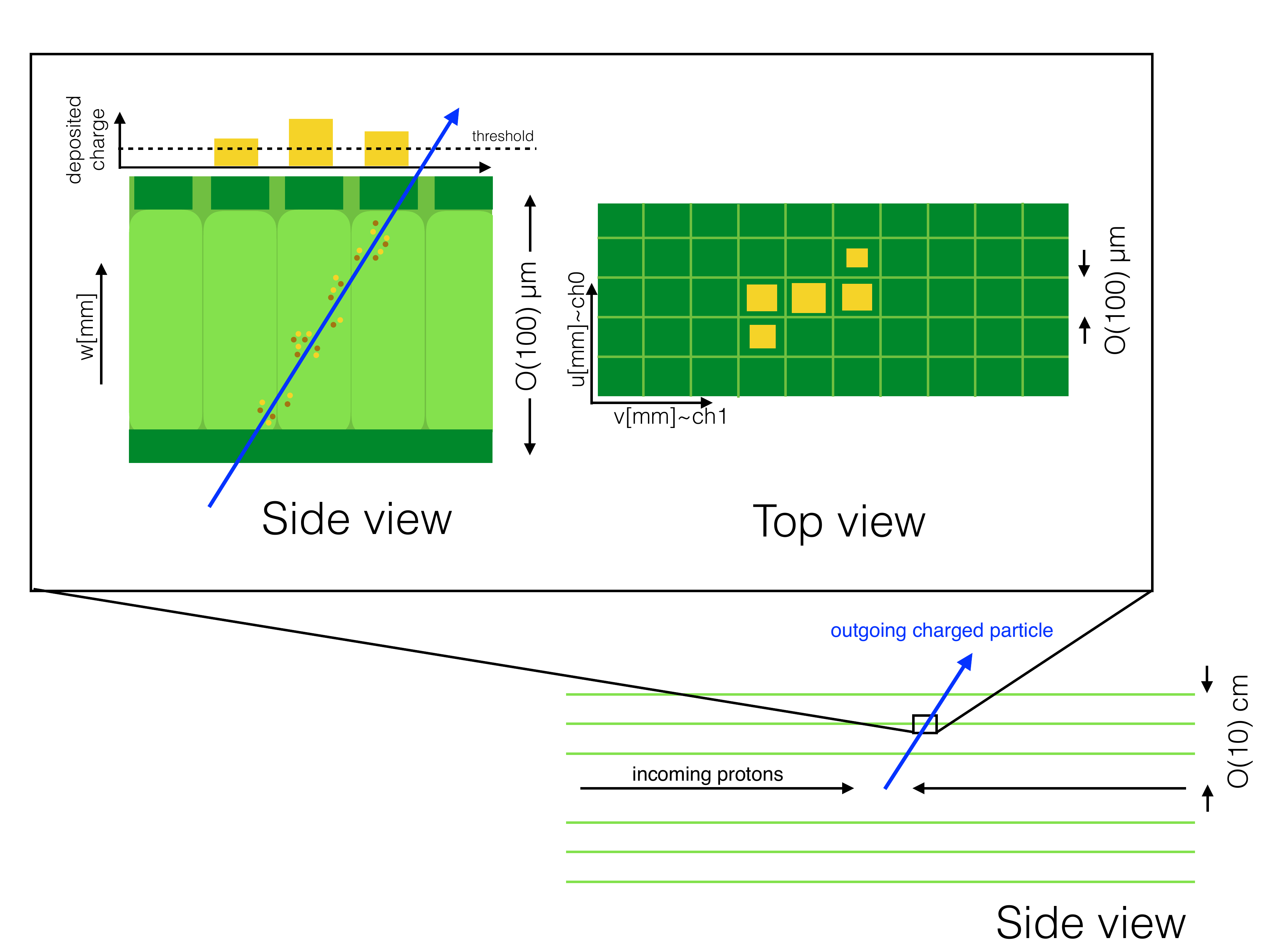}
    \caption{A schematic diagram of a proton-proton collision producing a charged particle that traverses layers of a pixelated tracking detector.  In reality, a given collision would result in many charged particles, but only one is indicated for illustration.
    \label{fig:schematic}}
\end{figure}

Other available measurements are the energy deposition and the structure of pixel clusters. In this paper we analyze the impact on seed purity of using cluster shape information.  The pixelation in tracking detectors is sufficiently fine that multiple pixels are often activated from a single charged particle.  A \textit{hit} is a pixel that has registered a signal above threshold and a collection of nearest neighbor hits are grouped together into a \textit{cluster}.  The shape of a pixel cluster contains useful information about the direction of the charged particle and offers a complementary source of information to the spatial location of the cluster centroid that is currently used for seed selection\footnote{Cluster shapes have been proposed for applications beyond centroid estimation in the forward region~\cite{Peilian,VIEL2016254}, but they can be useful more generally.}.  This is illustrated schematically in Fig.~\ref{fig:schematic}.  Thin layers of semiconductor provide the sensor material and with a strong applied electric field, ionized electrons and holes can be collected and registered with local readout electronics.  In some cases, these pixels may record a measurement of the deposited energy.  This information has a poor intrinsic resolution from straggling effects~\cite{Landau:1944if,Bichsel:1988if,Vavilov:1957zz} and is typically digitized with a small number of bits.  While this digitized information has many useful purposes (see e.g. Ref.~\cite{1710.02582}), the binary hit map carries the most important information and is the dominant source of additional information considered in this paper.

We propose to use neural networks to optimally combine local information about pixel cluster shapes with global geometric information to improve seed selection.  Neural networks have been recently studied for cluster seeding~\cite{Choma:2020cry,Dietrich:2019qif,DiFlorio:2018res}, but there has not yet been a systematic study of how much information local cluster structure adds to seed selection performance.  We compare classifiers trained to distinguish true seeds from noise seeds using local and/or global information.  Furthermore, we study neural networks applied to 1-, 2-, and 3-cluster seeds.  An $n$-cluster seed is a set of $n$ clusters that are used for constructing track candidates.  Even with a single cluster, one can use the information inside a pixel cluster to determine information about the underlying charged particle trajectory.

This paper is organized as follows.  Section~\ref{sec:simulation} introduces the TrackML~\cite{trackml,Amrouche:2019wmx} dataset, which we employ as a surrogate for actual LHC data, and defines the variables used in the analyses.  The results for one-, two-, and three-layer pixel clusters are presented in Sec.~\ref{sec:results}.  The paper ends with conclusions and outlook in Sec.~\ref{sec:concl}.

\section{Simulation}
\label{sec:simulation}

The TrackML dataset uses top quark pair production from proton-proton collisions as a representative process for track reconstruction at the LHC.  In order to emulate a realistic occupancy for the high-luminosity LHC, a Poission random number (with mean 200) of minimum bias events are overlaid on top of the $t\bar{t}$ collisions.  This leads to an average of about 10,000 particles/event. The hard-scatter and minimum bias events are both simulated using \textsc{Pythia}~\cite{Sjostrand:2006za,Sjostrand:2007gs}.  

The TrackML detector is a set of concentric cylindrical layers of pixelated sensors (i.e. pixel layers) complemented by a set of circular disks (i.e. strip layers) to ensure nearly $4\pi$ coverage in solid angle.  For the study presented in this paper, only the barrel pixel layers are used.  The pitch size of these pixel layers is 50 $\mu\text{m}$ in the direction perpendicular to the beam and 56.25 $\mu\text{m}$ in the beam direction.  The coordinates of clusters are determined as the charged-weighted average over the constituent hit locations.

Collisions occur near the geometric center of the simulated detector with Gaussian profiles that have mean zero and standard deviation 5.5 mm in the longitudinal direction (global $z$) and $15$ $\mu$m in the transverse directions (global $x$ and $y$).  The \textit{A Common Tracking Software} (\textsc{Acts}) toolkit~\cite{andreas_salzburger_2020_3741401} provides the simulation engine to propagate particles through a detector similar to ATLAS~\cite{Aad:2008zzm} or CMS~\cite{Chatrchyan:2008aa} at the LHC.  Particle trajectories are deterministic, except for multiple Coulomb scattering, ionizing energy loss, and radiation energy loss as emulated by the Fast Track Simulation Package in \textsc{Acts}.  Noise is simulated by  randomly adding additional hits (15\% of total hits).

\begin{figure}[t]
    \centering
    \includegraphics[width=0.45\textwidth]{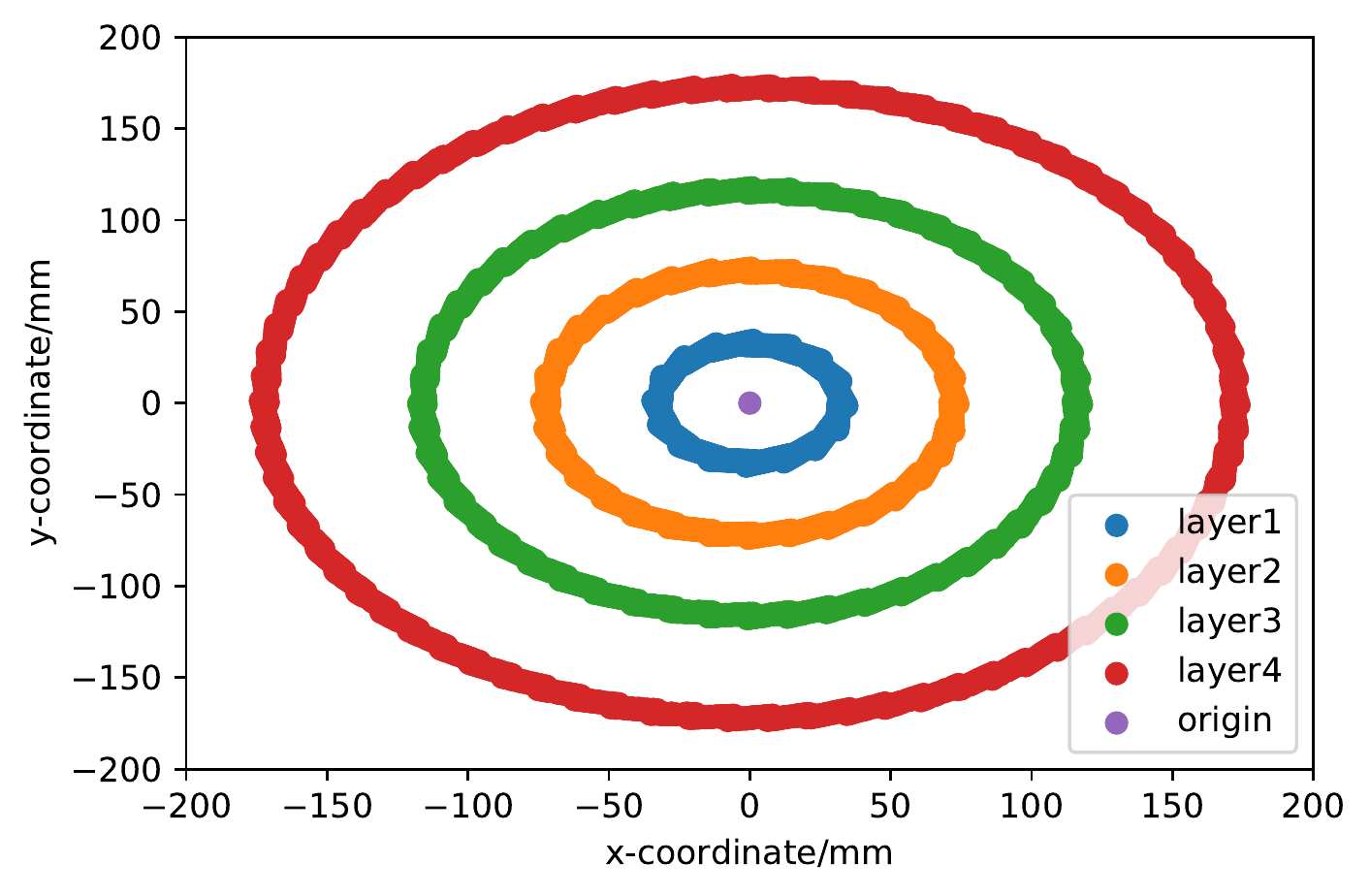} \includegraphics[width=0.45\textwidth]{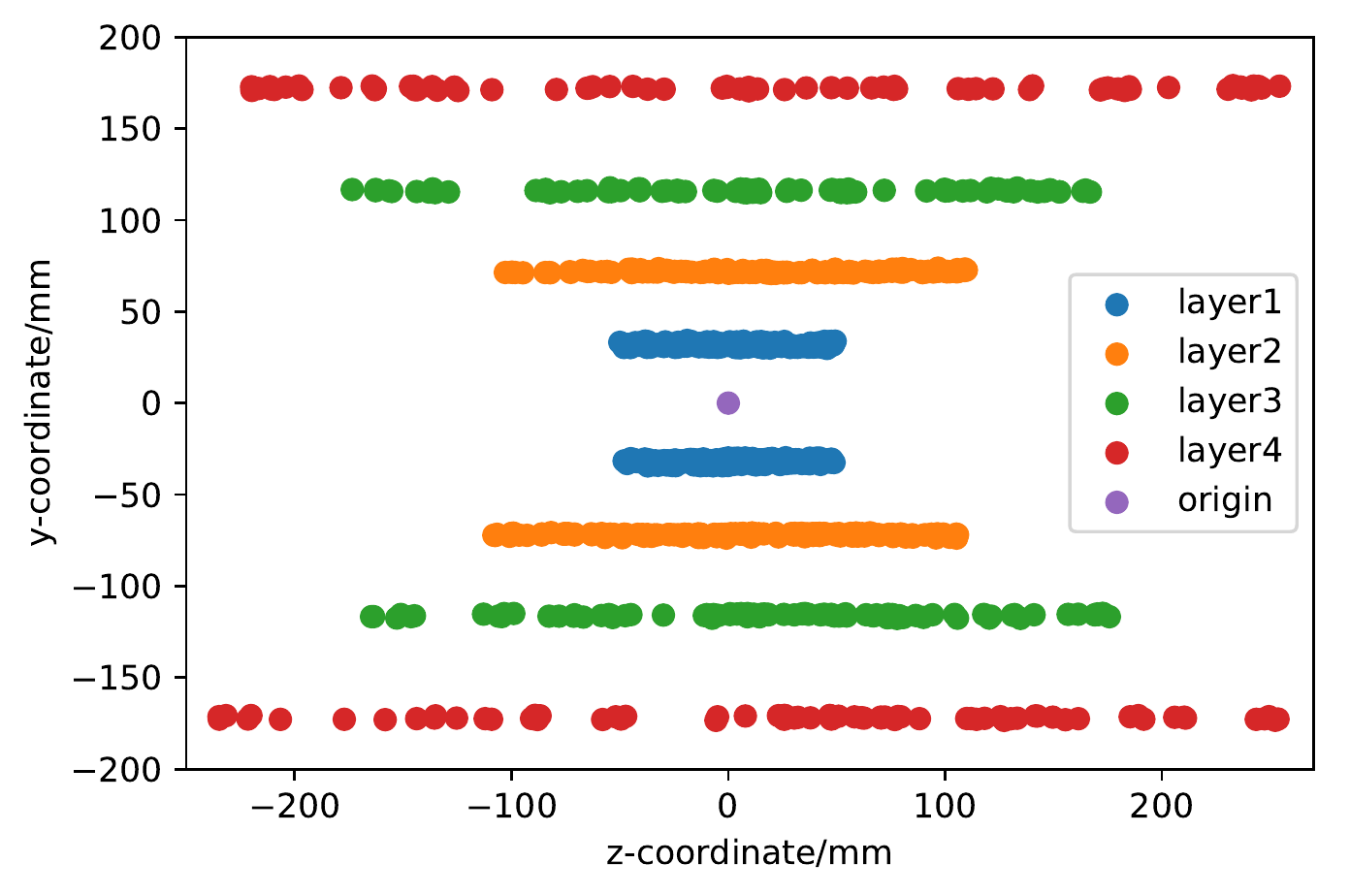}
    \caption{A sample of clusters in the inner most four barrel detector layers.  The $x$-$y$ view is on the left and the $y$-$z$ view is on the right.
    \label{fig:Scatter}}
\end{figure}

\begin{figure}[t]
    \centering
    \includegraphics[width=0.45\textwidth]{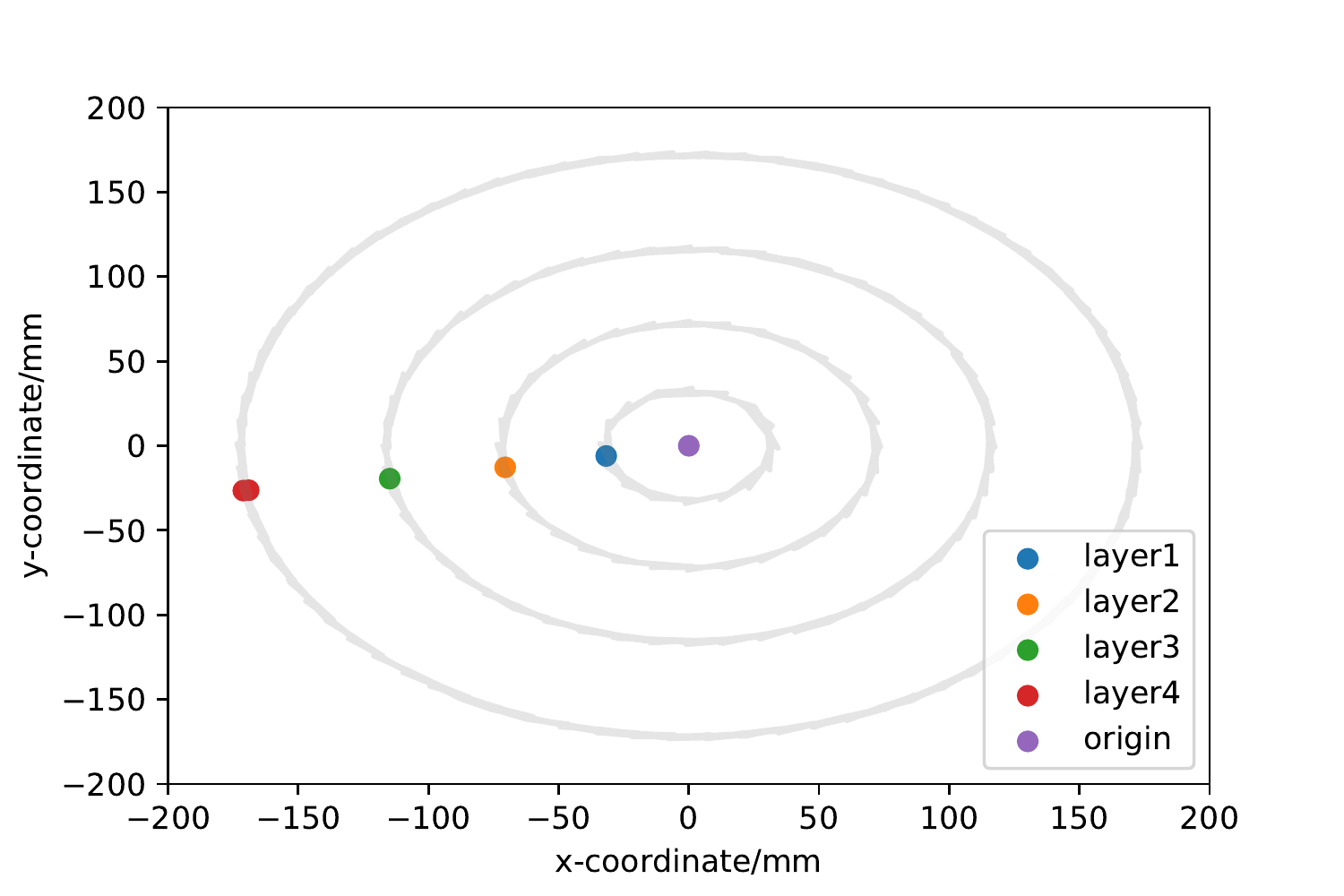}\includegraphics[width=0.45\textwidth]{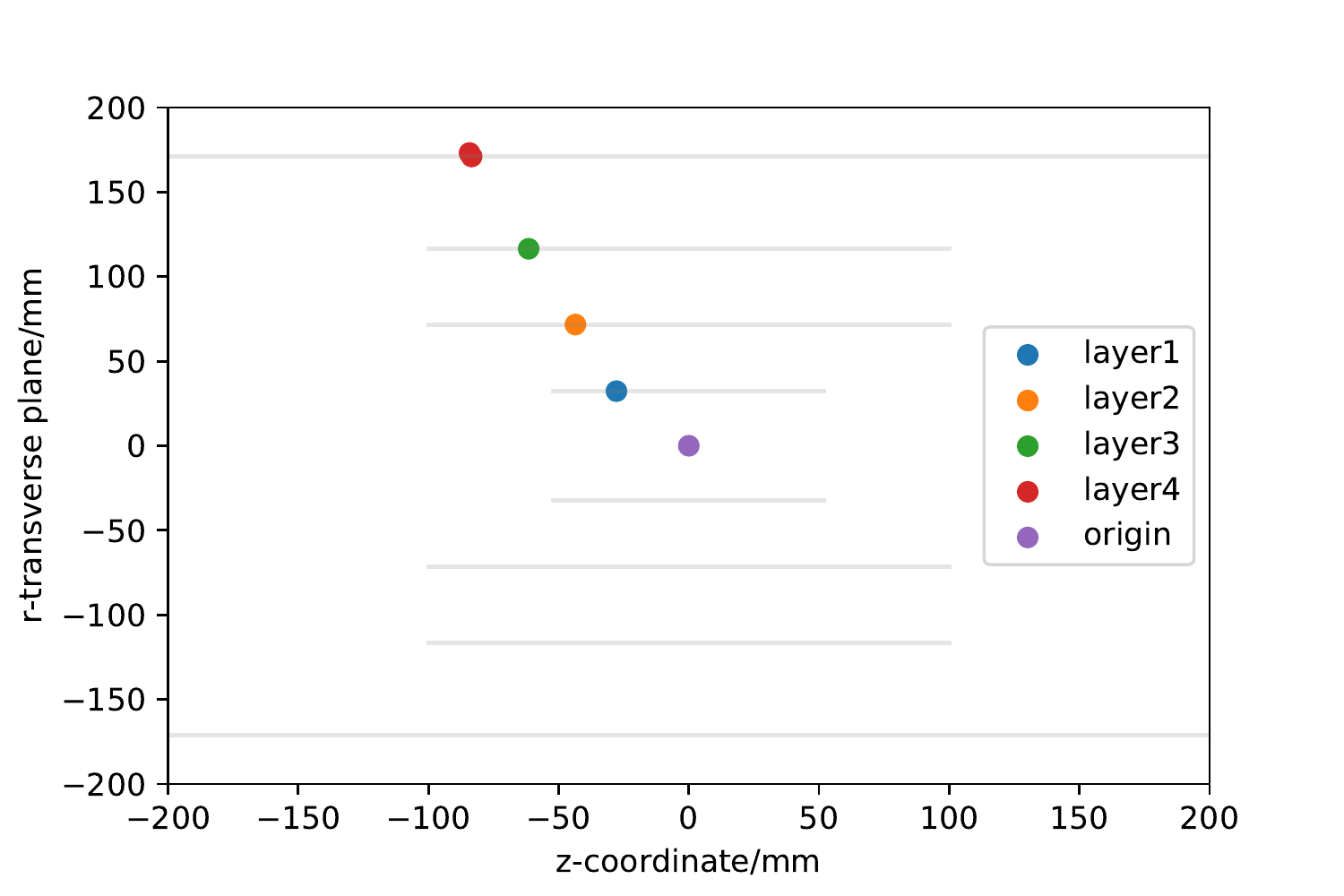}
    \caption{A representative example of pixel clusters from a single particle, in the $x$-$y$ plane (left) and the $z$-$r$ plane (right).
    \label{fig:Individual}}
\end{figure}

  A subset of clusters from one event are shown in Fig.~\ref{fig:Scatter} which illustrates the cylindrical nature of the inner barrel layers, and the clusters from a single particle are shown in Fig.~\ref{fig:Individual}.  Trajectories are approximately helical and manifest as arcs of circles in $x$-$y$ and straight lines in $r$-$z$.  Due to the small beamspot in $x$-$y$, the path in $x$-$y$ goes through the origin.  In contrast, there is a non-trivial spread in the beamspot in $z$; in the right plot of Fig.~\ref{fig:Individual}, the trajectory does not visually go through the origin.  %

In addition to the global position of the hit, there are the local coordinates that describe the location of the hit within a given module. These local coordinates are defined such that the $u$, and $v$ direction are along the active surface of the given module, with $u$ perpendicular to the beam direction, $v$ along the beam direction, and $w$ is the coordinate normal to the surface (see Fig.~\ref{fig:schematic}). The module defines the location of the hit using a pixel matrix, where the pixels in the $u$ direction are said to be in channel zero ($ch0$), and those in the $v$ direction are said to be in channel one ($ch1$). The number of pixel hits in $ch0$ and $ch1$ are denoted by $n_{ch0}$ and $n_{ch1}$ respectively. 
The cluster lengths in the $u$ direction ($\Delta u$) and in the $v$ direction ($\Delta v$) are given by multiplying the pitch sizes of the cells by $n_{ch0}$ and $n_{ch1}$ respectively.  Therefore, the cluster shape is defined by [$\Delta u$, $\Delta v$, $\Delta w$], where $\Delta w$ is the width of the module.

\begin{figure}[t]
    \centering
    \includegraphics[width=0.65\textwidth]{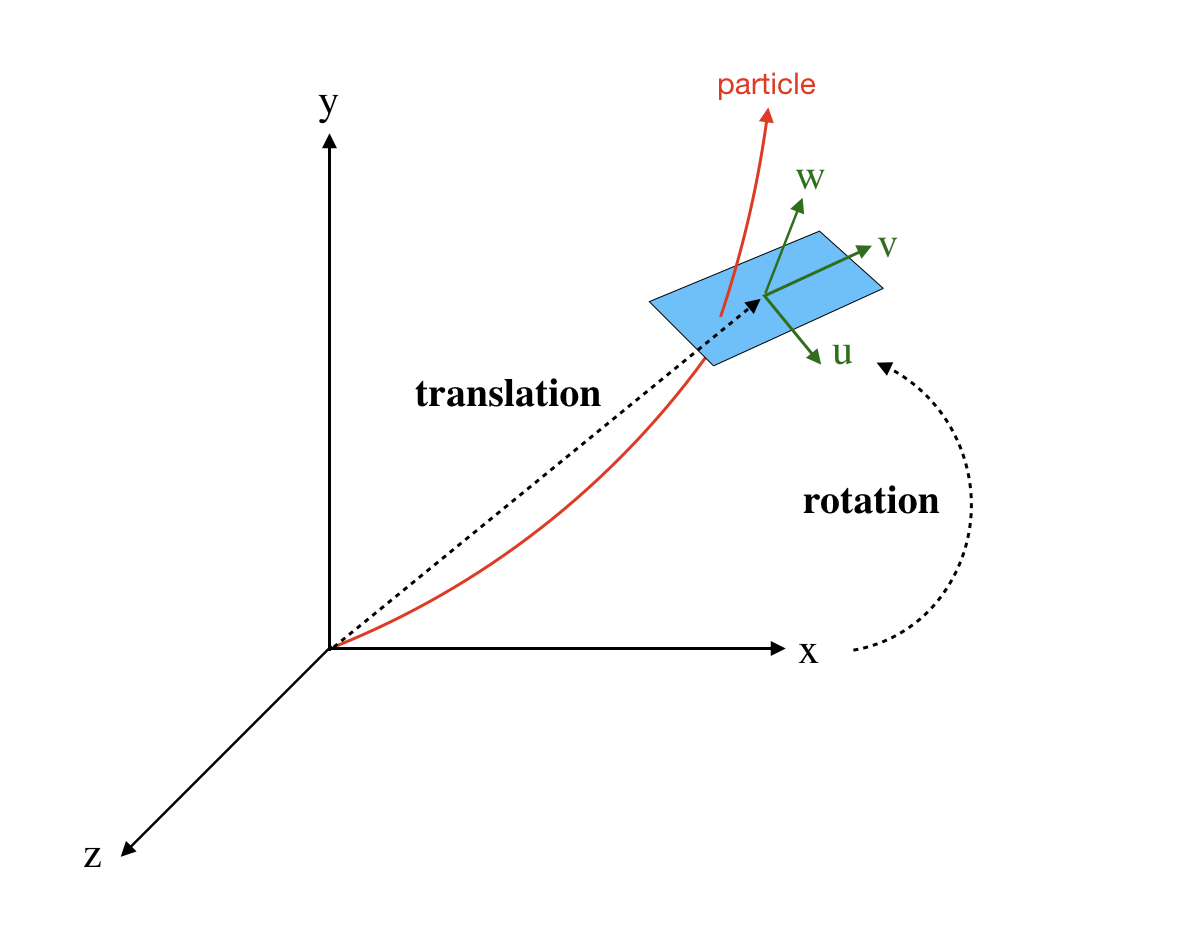}
    \caption{Example of transformation from local coordinates to global coordinates. Reproduced from the TrackML challenge webpage~\cite{trackml}.}
    \label{fig:localToGlobal}
\end{figure}

In many instances, more than one pixel will have charge deposited within it. In order to reconstruct the hit location, a weighted average of the charge deposited information is used, i.e.:
\begin{equation}
    \vec{h}_{\mathrm{local}} = \frac{1}{\sum_i q_i}\sum_i q_i\, \vec{p}_i,
\end{equation}
where the sum goes over all pixels with deposited charge above a threshold, $q_i$ is the charge deposited, and $\vec{p}_i = (u_i,\,v_i,\, w_i)$ is the local position of the pixel in millimeters. After reconstructing the local hit position information, the hit position is translated to global coordinates within the detector. This is achieved through the use of a rotation matrix and a translation vector:
\begin{equation}
\label{eq:translocaltoglobal}
    \vec{h}_{\mathrm{global}} = U_{\text{rot}}\,\vec{h}_{\mathrm{local}} + \vec{x}_{\text{trans}},
\end{equation}
where both the translation vector, $\vec{x}_{\text{trans}}$, and rotation matrix, $U_{\text{rot}}$, are module dependent as illustrated in Fig.~\ref{fig:localToGlobal}.  In what follows we make use of $\vec{h}_{\mathrm{local}}$,  $\vec{h}_{\mathrm{global}}$ and $\vec{p}_i$.

\section{Results}
\label{sec:results}

The following section explores the information available from a single (Sec.~\ref{sec:onecluster}), from a pair (Sec.~\ref{sec:twoclusters}), and from a triplet (Sec.~\ref{sec:threeclusters}) of pixel cluster(s).  To illustrate the potential gains from local cluster shape information, we will use deep neural networks trained using three feature sets. First, we will train neural networks to only consider the spatial information contained in the global $x$,$y$,$z$ coordinate per pixel cluster (based on the cluster centroid).  Next, we consider networks trained using only local information about the pixel cluster shape.  Finally, both sets of features are combined to demonstrate the optimal combination of both local and global information.  As will be shown below, the shape of the cluster can help to narrow down the search region by providing pointing information on where the next hit in the subsequent layer should be.  The similarity of cluster shapes between layers when considering pairs or triplets can also reject incorrect pairings.  By using this information, the number of seeds that need to be considered can be reduced, improving subsequent steps in the seeding algorithm.

All neural networks are multilayer perceptrons with fully connected layers using the rectified linear unit (ReLU) activiation function between layers.  These networks are optimized by the gradient-based stochastic algorithm Adam~\cite{adam}.

\subsection{Single clusters}
\label{sec:onecluster}

For a single pixel cluster, the networks are trained to predict the direction of the particle in the frame defined by the local detector unit. We found that choosing to use the momentum defined in the module reference frame instead of attempting to predict the direction in a globally defined coordinate system improves the accuracy of the network's predictions.  The neural networks consist of five layers using $[256,128,64,32]$ units in the hidden layers. The final output is a 3-vector prediction for the particle momentum, $\vec{p}_\text{pred}$, at the location of the hit in the frame defined locally by the module. Here, it is safe to neglect the effect of the magnetic field since we are looking at the inner most layers which are close enough together that the impact of the magnetic field from one layer to the next does not modify the momentum in a significant manner. The networks are trained with a cosine-proximity loss function, which minimizes the angle, $\Delta\theta$,  between the true momentum $\vec{p}_\text{true}$ and the predicted momentum $\vec{p}_\text{pred}$,
\begin{equation}
\cos\Delta\theta = \frac{\vec{p}_\text{pred} \boldsymbol{\cdot} \vec{p}_\text{true}}{|\vec{p}_\text{pred}| |\vec{p}_\text{true}|}~.
\end{equation}
Thus, the normalisation of the network's output is unimportant.  It would be interesting to see if the network could be trained to predict the direction and magnitude of the momentum, but this is a more difficult task.  Since the change in momentum is small between consecutive hits in the inner layers this additional information could be used to further reduce the combinatorics when constructing doublets.  

The dimension of the input vector to the neural networks depends upon which set of features are being considered.  There are three inputs for the spatial information ($x$-$y$-$z$) and eight inputs associated with the cluster shape.  In reality, a cluster may have charge deposits in more than 50 pixels, or as few as one.  However, to encode the shape of the cluster in fewer variables we first find the convex hull of the hits in the cluster using the Graham scan algorithm~\cite{GRAHAM1972132}. We then define a bounding box around the convex hull by identifying the two extremum pixels. The eight input variables for the cluster are then given as the position of these two extremum pixels (four variables), the charge deposited in these pixels (two variables), and the cluster shape ($\Delta u$, $\Delta v$). Although this information contains redundancy we found it was beneficial with respect to training.

\begin{figure}[t]
    \centering
    \includegraphics[width=0.75\textwidth]{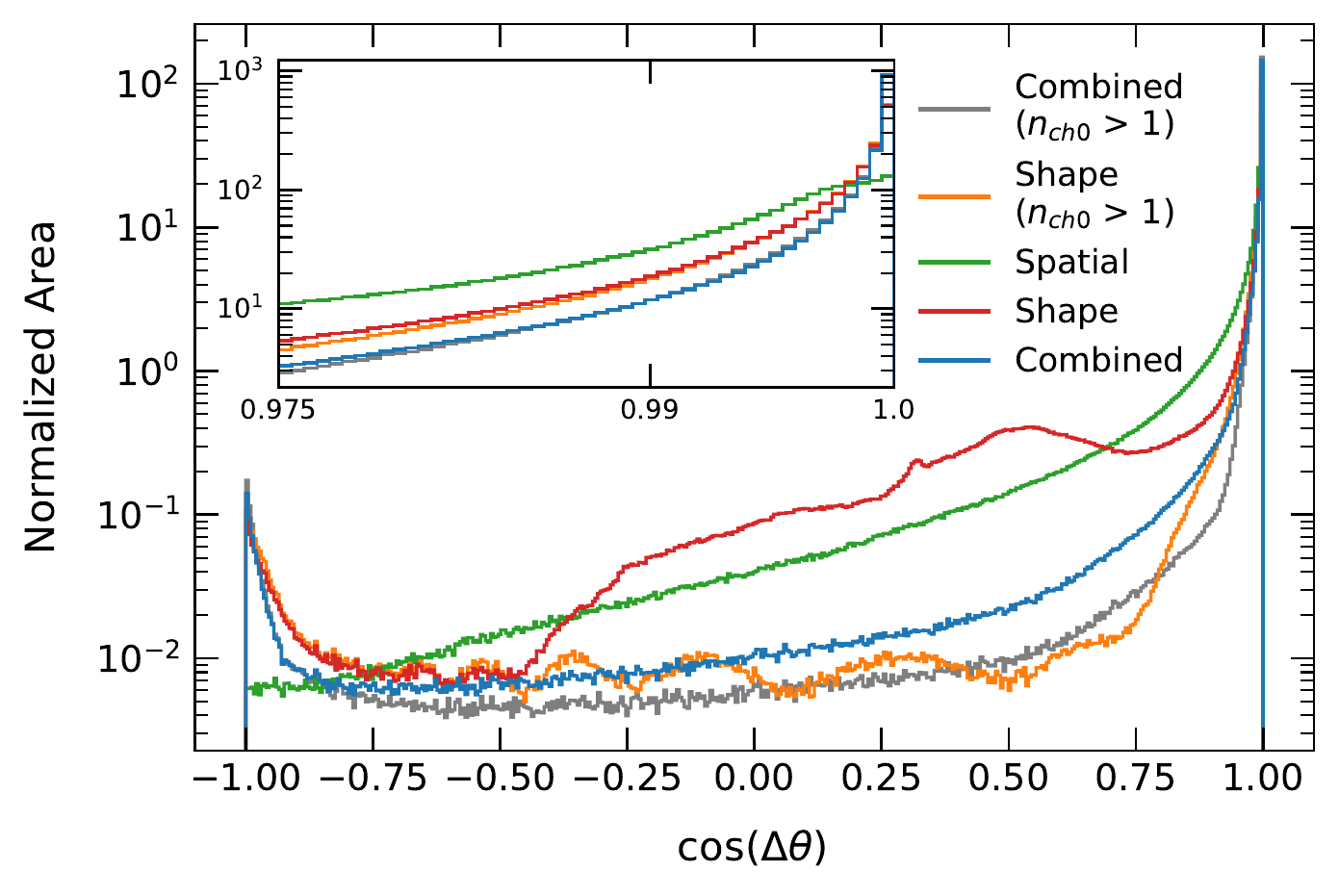}
    \caption{The number of counts versus the cosine between the true direction and the predicted direction for hits in the tracker, using spatial (green), shape (red), and combined (blue) information. Furthermore, if the cluster shape is required to have a width in channel 0 greater than 1, the results for the shape and combined are updated to the orange and gray lines respectively.}
    \label{fig:singlet_cosine}
\end{figure}

After the training is complete, the networks are compared using three different metrics.
The first metric is to compare the cosine of the angle between the true direction of the particle and the predicted direction from the network. Since we are only considering the inner most layers for this network, the subsequent hit should be further from the collision point than the initial. Therefore, if the network predicts the momentum to be towards the center, we can flip the direction without introducing additional errors. The different networks are compared in Fig.~\ref{fig:singlet_cosine} for this metric. While the shape-only network has a sharper peak at $\cos(\Delta\theta) = 1$ than the spatial-only network, the shape-only network has a bump at $\cos(\Delta\theta) \in [-0.25, 0.75]$ degrading the overall performance of the shape-only network in comparison to the spatial-only network. Furthermore, by combining the spatial and the shape information, there is a significant improvement in the ability to predict the true direction of the particle. The most interesting aspect is related to  clusters in which the track does not fall within a single value of channel zero (row) on the detector (see Fig.~\ref{fig:schematic}). In the case of only using shape information, when the cluster falls along a single row, the network is unable to predict in which direction the hit went through the layer. This introduces the bump in the prediction for the shape information in Fig.~\ref{fig:singlet_cosine} due to the degeneracy over possible angles consistent with the width and breadth of the cluster shape. Giving the network access to the information about the global position of the hit allows it to determine the most likely direction of the particle. This results in the improvement seen in the combined prediction. However, if we remove the single row cluster shapes (which make up approximately 30\% of all hits), then the network's performance is improved even further as shown by the gray (combined) and orange (shape) curves.

\begin{figure}[t]
    \centering
    \includegraphics[width=0.75\textwidth]{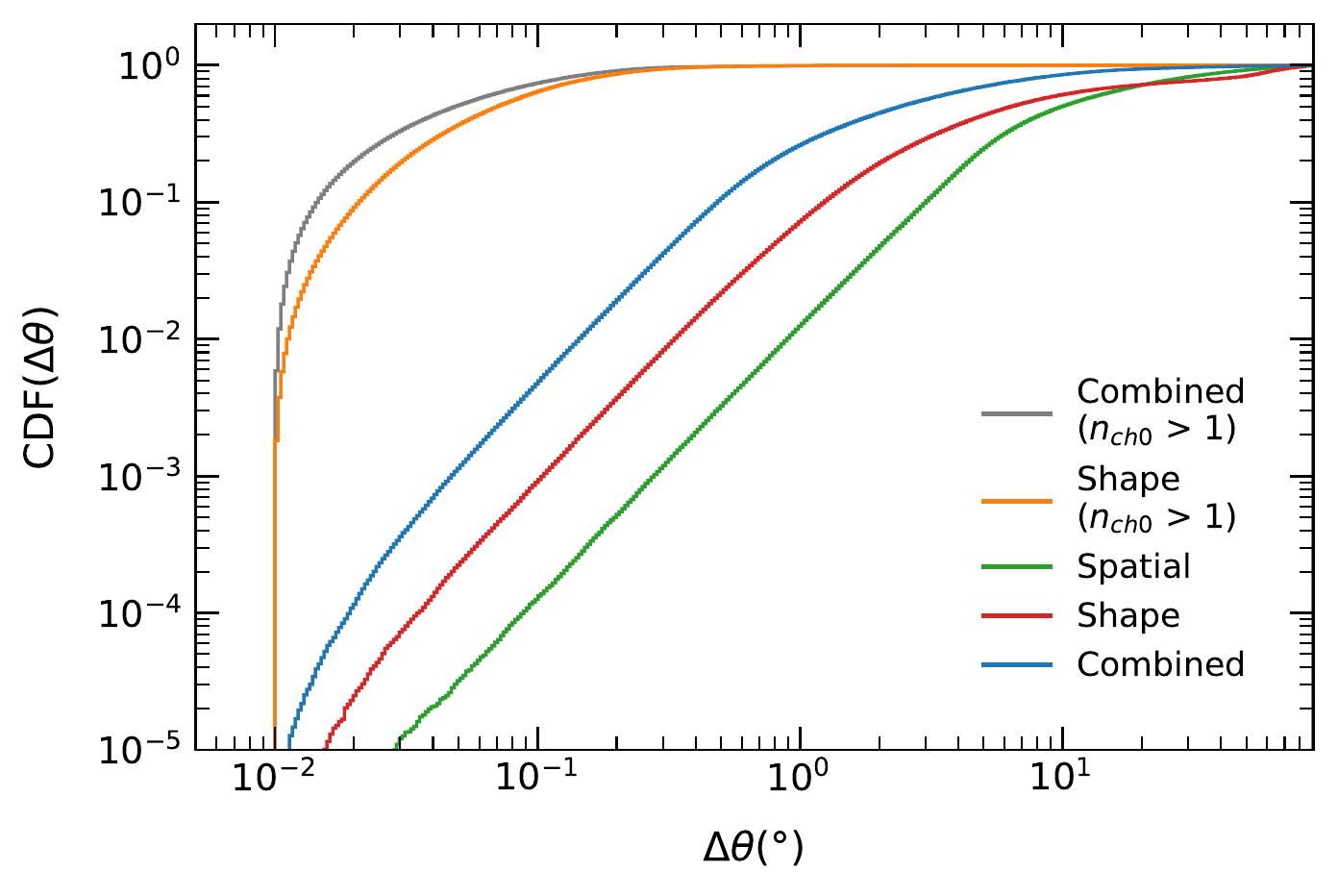}
    \caption{The fraction of events with a predicted direction less than $\Delta\theta$ away from the true direction. The results are shown using the spatial information only (green), the shape information only (red), and the combined (blue). Furthermore, if the cluster shape is required to have a width in channel 0 greater than 1, the results for the shape and combined are updated to the orange and gray lines respectively.}
    \label{fig:singlet_cdf}
\end{figure}

The second metric is to consider the fraction of events for which the predicted direction falls within a cone of a given angle from the true direction. The smaller the cone required the better, since this would help to reduce the number of combinations to consider during the seeding process. The results are shown in Fig.~\ref{fig:singlet_cdf}. Again, the shape-only network has more predictions close to the true direction, but has a very long tail as compared to the spatial information. By combining the two however, significant improvement is made by both having approximately 50\% of all predicted directions within $2\degree$ of the true direction, but by also removing the long tail. In comparison, the spatial only does not reach a 50\% containment until almost $10\degree$ and the shape only does not reach 50\% until about $5\degree$. In the case of removing single row cluster shapes, we can see that the network is able to predict all of the hit directions to within $2\degree$ of the true direction.

\begin{figure}[t]
    \centering
    \includegraphics[width=0.45\textwidth]{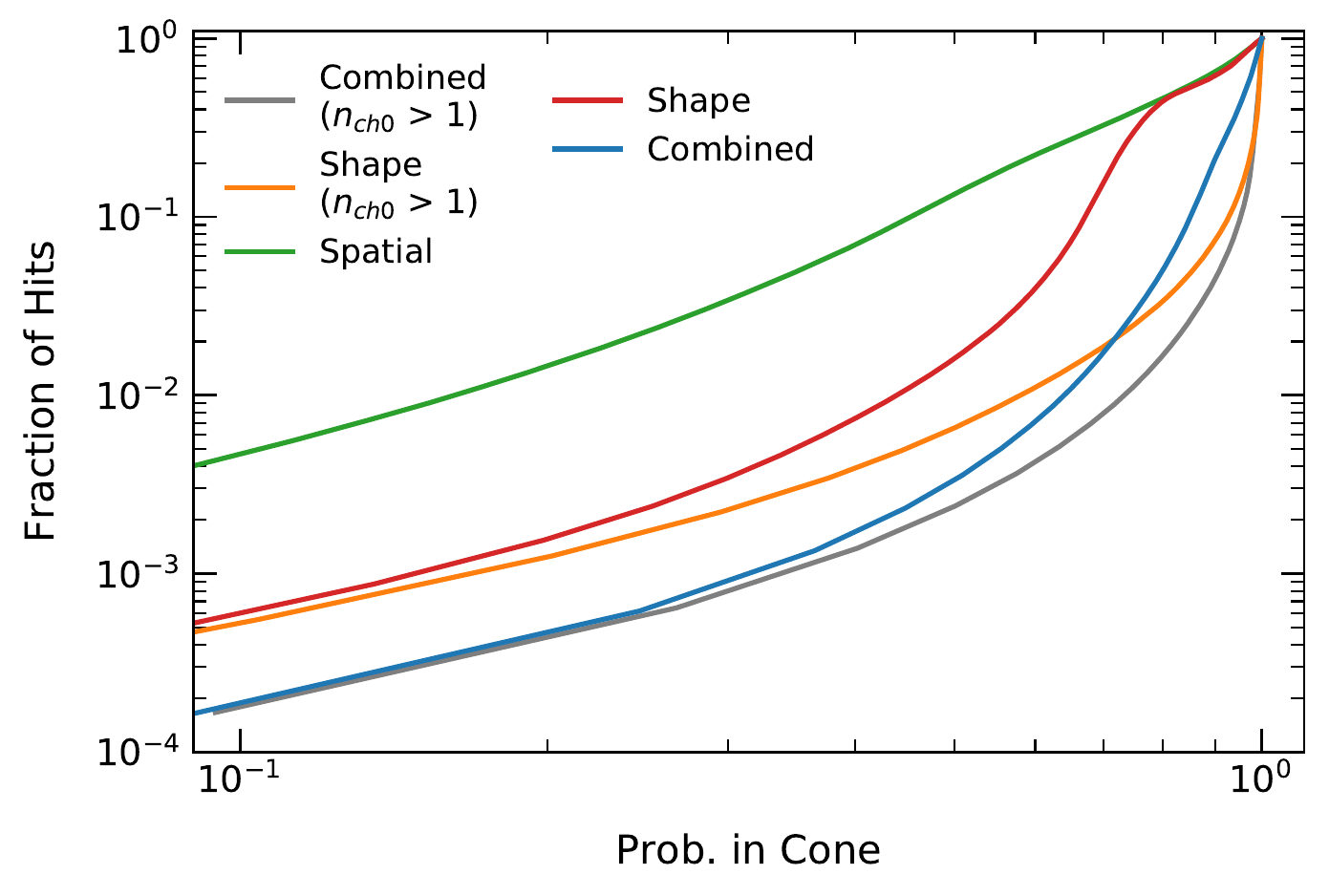}
    \includegraphics[width=0.45\textwidth]{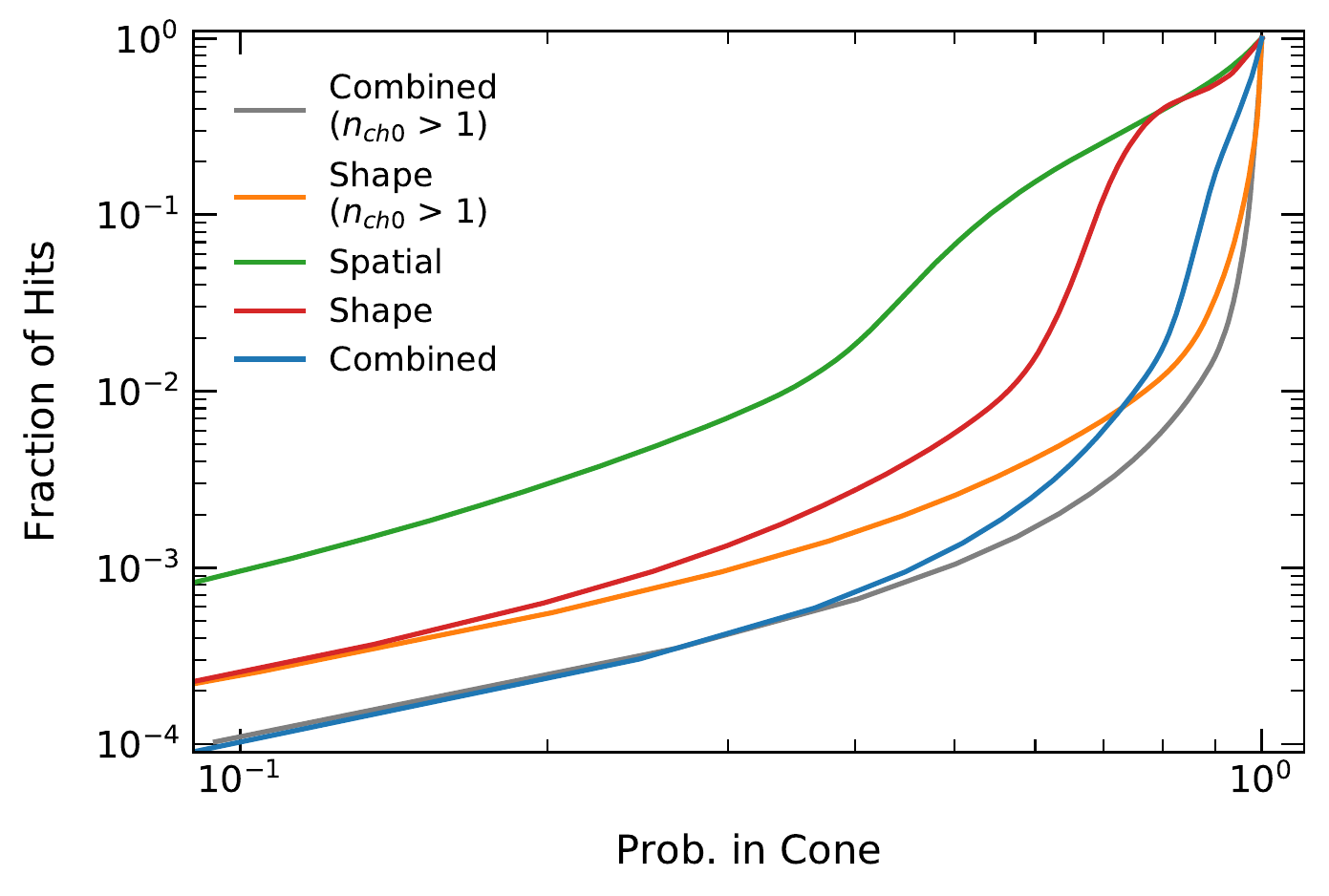}
    \caption{The fraction of hits in the next layer that fall within a cone are shown against the probability that the true hit is contained within that cone. This was obtained by averaging over all track hits in layers 2 and 4 from 100 events. The curves are parameterized by the opening angle of the cone. The results for spatial information only (green), shape information only (red), and combined (blue/gray) are shown considering all hits (left) and taking into account that the hit must be moving away from the collision point (right). The results shown here are for the innermost layer to the next layer. Requiring that the width in channel 0 is greater than 1 moves the combined curve to the gray curve shown, and moves the shape only to the orange curve.}
    \label{fig:singlet_roc}
\end{figure}

The final metric is the comparison of the probability that the true hit is within a cone of a given size compared to the fraction of all the hits in the event that lie inside the cone. The angular opening of the cone parameterizes the curve defining this metric. The results are shown in Fig.~\ref{fig:singlet_roc}, which was made by considering 100 events each with about 1000 hits  in the innermost layer and looking only at the hits in the innermost layer to the next layer out. The left plot shows the results if the outward direction of the hit from the interaction point is not taken into account, while the right plot takes the direction into account. If we take a working point with 95\% probability of finding the true hit inside the cone and considering that the hit has to be moving outwards, then we have approximately a 38.5\% of all hits within the cone for the combined network, which improves to 4.60\% when only cluster shapes with a channel 0 width greater than 1 are considered. The accuracy degrades slightly when neglecting the outward direction to 41.1\% and 9.4\% respectively. The effect of using wider clusters is only important in the region of high probability of finding the hit within the cone. If instead of a 95\% chance to find the hit we use a 50\% chance, then the fraction of hits within the cone is 0.13\% and 0.1\% for the combined and combined with channel 0 greater than 1 respectively.  If this network is used to produce a triplet seed, we expect the reduction in possible seeds to scale as the fraction of hits in the cone squared.  Table~\ref{tab:prob_cone} shows the fraction of hits within the cone for all results at a 50\% and 95\% probability of finding the hit in the cone.

\begin{table}[ht]
    \centering
    \begin{tabular}{ccc}
    \hline\hline
    & \multicolumn{2}{c}{Fraction of hits in cone} \\
    Network & 50\% Prob. Working Point & 95\% Prob. Working Point\\
    \hline
    Spatial & 0.07 & 0.75 \\
    Cluster & 0.006 & 0.70 \\
    Combined & 0.001 & 0.39 \\
    Cluster ($n_{ch0} > 1$) & 0.003 & 0.09 \\
    Combined ($n_{ch0} > 1$) & 0.001 & 0.05 \\
    \hline\hline
    \end{tabular}
    \caption{Efficiency and purity numbers at two possible working points, based on Figure~\ref{fig:singlet_roc}.}
    \label{tab:prob_cone}
\end{table}


We have not attempted to balance the trade-off between efficiency ($\epsilon$) and purity to find an optimal working point for the single cluster NN approach but instead conclude this section with a few observations.  One possibility could be to work at a lower efficiency value, which results in a  smaller number of background hits in the cone, but to apply the direction prediction on consecutive layers and compare them.  
Possibly allowing working with lower efficiency but taking into account both forward and backward predictions.   Alternatively, the single hit network could be used to predict the particle directions for all hits within consecutive layers and compare these directions to find mutually consistent sets. 

For hit patterns involving more than a single line of hit pixels the combined network is remarkably adept at correctly predicting the particle's true direction, which potentially motivates training specialized networks for different classes of pixel clusters. One such specialized network would be the Bayesian Neural Network~\cite{doi:10.1162/neco.1992.4.3.448}, in which the network would be trained to predict both the track direction and an uncertainty on that direction.  Even for the single row hit patterns there is additional information, which we have not utilised, in the fraction of charge deposited in each pixel.  Finally, the output from this single hit network can be fed to, or used in tandem with, another network to determine which of the possible hits in the cone are true track partners of the original hit, this will be addressed in the subsequent sections.

\subsection{Cluster Doublets}
\label{sec:twoclusters}
\begin{figure}[htb]
    \centering
    \includegraphics[width=0.65\textwidth]{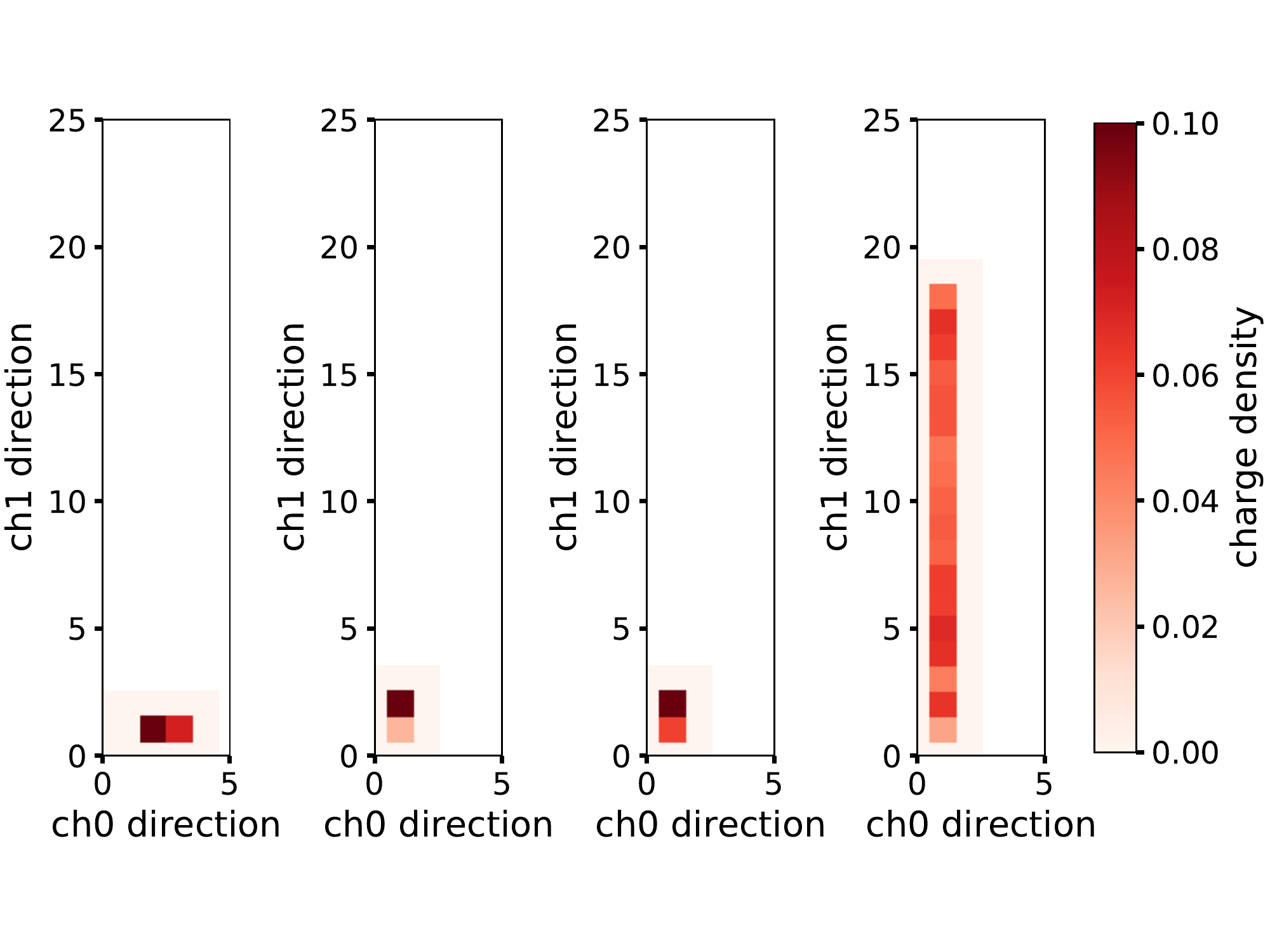} 
    \caption{Cluster shape of two clusters coming from the same track (left two) and two clusters coming from different tracks (right two) for two clusters forming a doublet with $\pt < 0.5$~GeV and $|\eta| < 1$ . Note that the origin (bottom left) of $ch0$ and $ch1$ has been shifted to the position where the cluster is located, for clarity. The pixel intensity demonstrates the charge density deposited in the pixel, but this information is not given to the NN.}
    \label{fig:cluster_low}
\end{figure}

\begin{figure}[htb]
    \centering
    \includegraphics[width=0.65\textwidth]{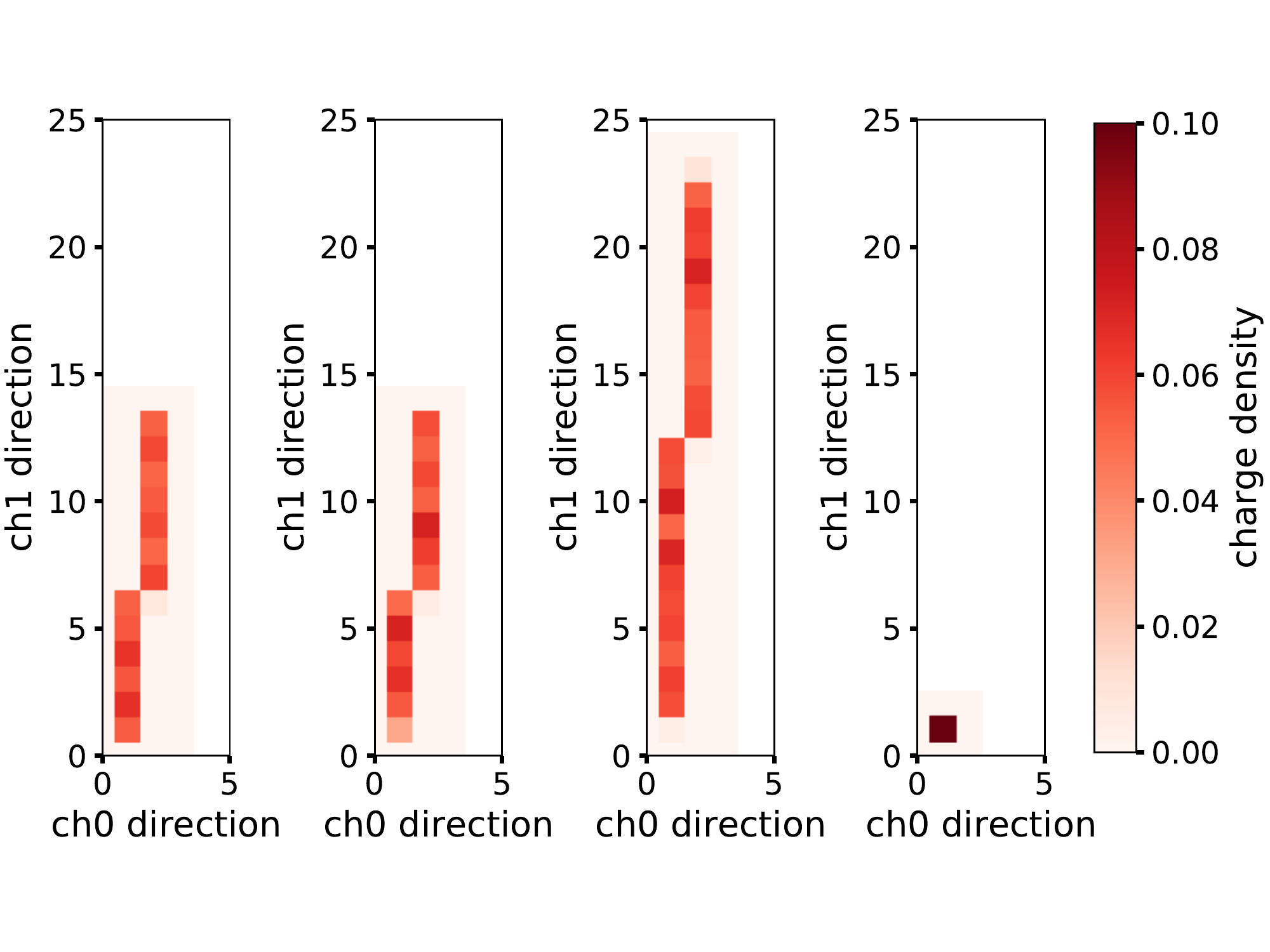} 
    \caption{Cluster shape of two clusters coming from the same track (left two) and two clusters coming from different tracks (right two) for two clusters forming a doublet with $\pt > 1$~GeV and $2.1 < |\eta| < 3$ . Note that the origin (bottom left) of $ch0$ and $ch1$ has been shifted to the position where the cluster is located, for clarity. The pixel intensity demonstrates the charge density deposited in the pixel, but this information is not given to the NN.}
    \label{fig:cluster_hi}
\end{figure}

Doublets are composed of two clusters coming from consecutive layers. Constructing doublets is the first step towards reconstructing a track, which is a crucial step not only in  traditional track reconstruction algorithms~\cite{Aaboud:2017all} but also in deep learning based algorithms~\cite{Farrell:2018cjr,Ju:2020xty,choma2020track}. %
We now discuss our doublet neural network, which was trained to separate true doublets from fake doublets. The list of possible doublets may come from all possible pairs of clusters in consecutive layers, or may be from the output of the direction prediction discussed in the previous section.   Doublets formed by clusters coming from the same charged particle are true doublets; those formed otherwise are fake doublets.  We define the purity as the ratio of true doublets to total doublets in the given selection. Similarly, the efficiency is defined as the ratio of the number of true doublets that pass selection requirements to the total number of true doublets input to the selection process. We demonstrate that the neural network trained with the combined information of cluster spatial and shape information results in a higher doublet purity for a given doublet efficiency.  The reason for this is that the cluster shape of two clusters coming from the same track tend to have similar structures. As an illustration, Figure~\ref{fig:cluster_low} and \ref{fig:cluster_hi} show the cluster shapes of a pair of clusters coming from the same track (left two panels) and a pair of clusters coming from different tracks (right two panels) in two different kinematic regions. 

The doublet neural network is composed of four layers of neurons with the sizes of [128, 64, 32, 1]. The outputs of the last neuron are  transformed by a sigmoid function and are used as the doublet scores: the network's prediction for the probability that the doublet is a true doublet. To balance the number of true and fake doublets in the training, true doublets are constructed from 100 events and fake doublets are from a single event using all possible combinations of the hits in the two innermost consecutive layers. The doublet scores are quantitatively compared with the labeling by using the binary cross-entropy loss function. The smaller the loss value is, the better the predictions agree with the ground truths.

\begin{table}[htb]
    \centering
        \caption{Input variables for training the doublet neural network. The definition of the variables can be found in the text. %
        \label{tab:doublet_inputs}}

    \begin{tabular}{cc}
    \hline\hline
        Group & Variables \\ \hline
        spatial only & $\Delta r$, $\Delta\phi$, $\Delta\phi / \Delta r$, $\Delta z$, $z_0$, $\Delta\eta$ \\ 
        shape only & $\Delta\eta^\prime$, $\Delta\phi^\prime$ \\  \hline\hline
    \end{tabular}
\end{table}

The doublet neural network is trained separately with the same training events but different sets of input variables as summarized in Table~\ref{tab:doublet_inputs}. `Spatial only' variables are derived from the cluster's global position in a cylindrical coordinate system. $\Delta r$, $\Delta\phi$, $\Delta z$ are the differences of the two hits in  $r$, $\phi$ and $z$, $\Delta\eta$ is the difference in pseudorapidity\footnote{Pseudorapidity $\eta = -\ln(\tan{(\theta/2)})$, where $\theta$ is the polar angle relative to the beamline.} and $z_0$ is the intercept of the line connecting the two hits at $r=0$. `Shape only' variables are derived from the cluster shape described in Section~\ref{sec:simulation}.  The length of the cluster is given by $L=(\Delta u^2+\Delta v^2+\Delta w^2)^{1/2}$.  We then use the rotation matrix in Eq.~(\ref{eq:translocaltoglobal})
to transform the cluster shape to a  coordinate system aligned with the global coordinates, in which the shape is [$\Delta u'$, $\Delta v'$, $\Delta w'$]. In the new coordinate system, 
we introduce two new variables, $\eta'$ and $\phi'$, that are related to the fraction of the length of the cluster along the beam axis and its orientation in the plane perpendicular to the beam, respectively:
\begin{equation}
\eta'= \frac{1}{2}\log\frac{L-\Delta w'}{L+\Delta w'}~,\quad\quad \phi'=\tan^{-1}\frac{\Delta v'}{\Delta u'}~.
\end{equation}
The quantities $\eta^\prime$ and $\phi^\prime$ are calculated for each cluster and the differences  between the two clusters, $\Delta\eta'$, $\Delta\phi'$, are used as inputs to the network. Redundant spatial information is fed to the network to gain accuracy and convergence time. 

\begin{figure}
    \centering
    \includegraphics[width=0.5\textwidth]{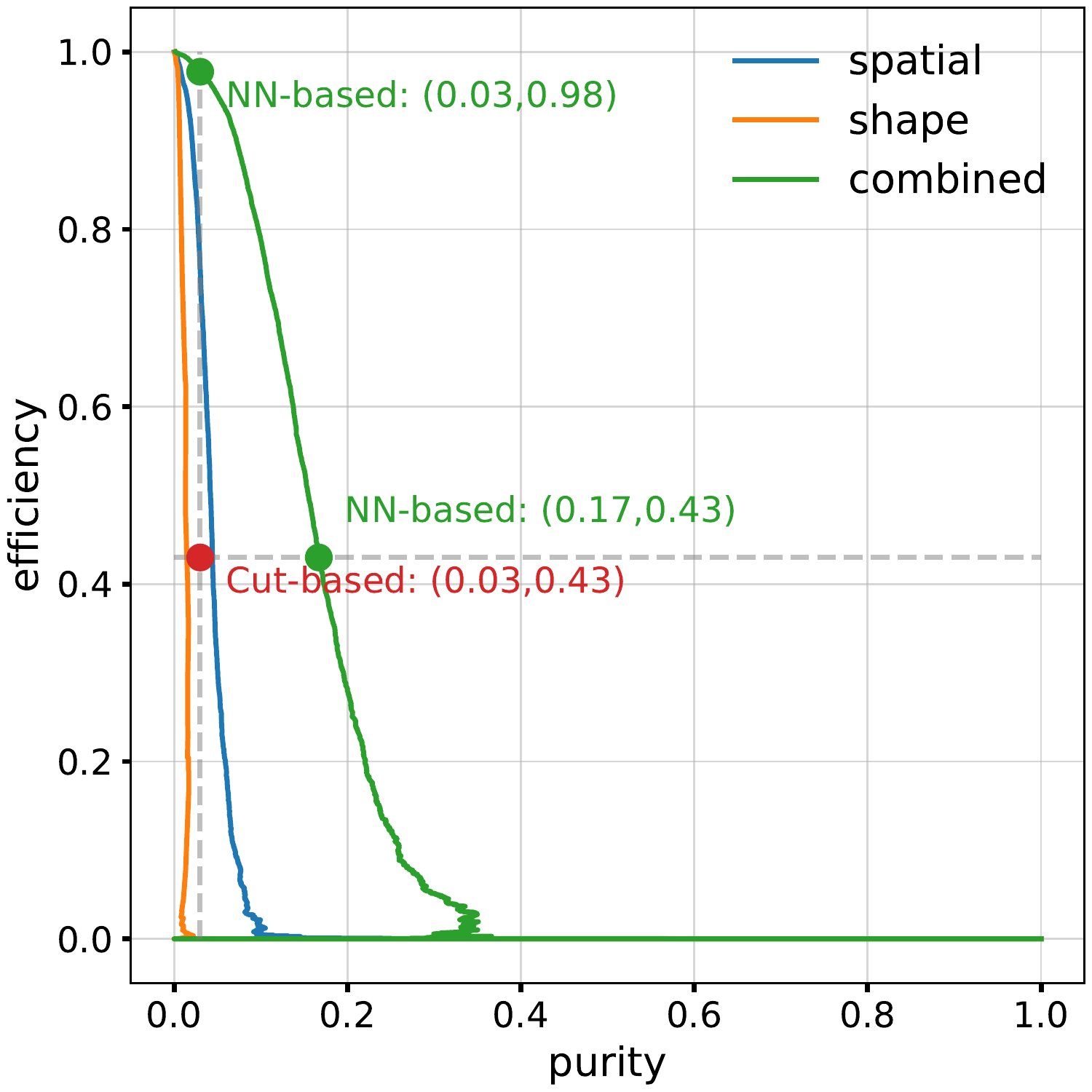}
    \caption{Doublet efficiency versus doublet purity for different trained doublet neural networks and the cut-based doublet selection. The definition of doublet efficiency and purity can be found in the text.
    }
    \label{fig:doublet_vs_cutbased}
\end{figure}

\begin{figure}
    \centering
    \includegraphics[width=0.8\textwidth]{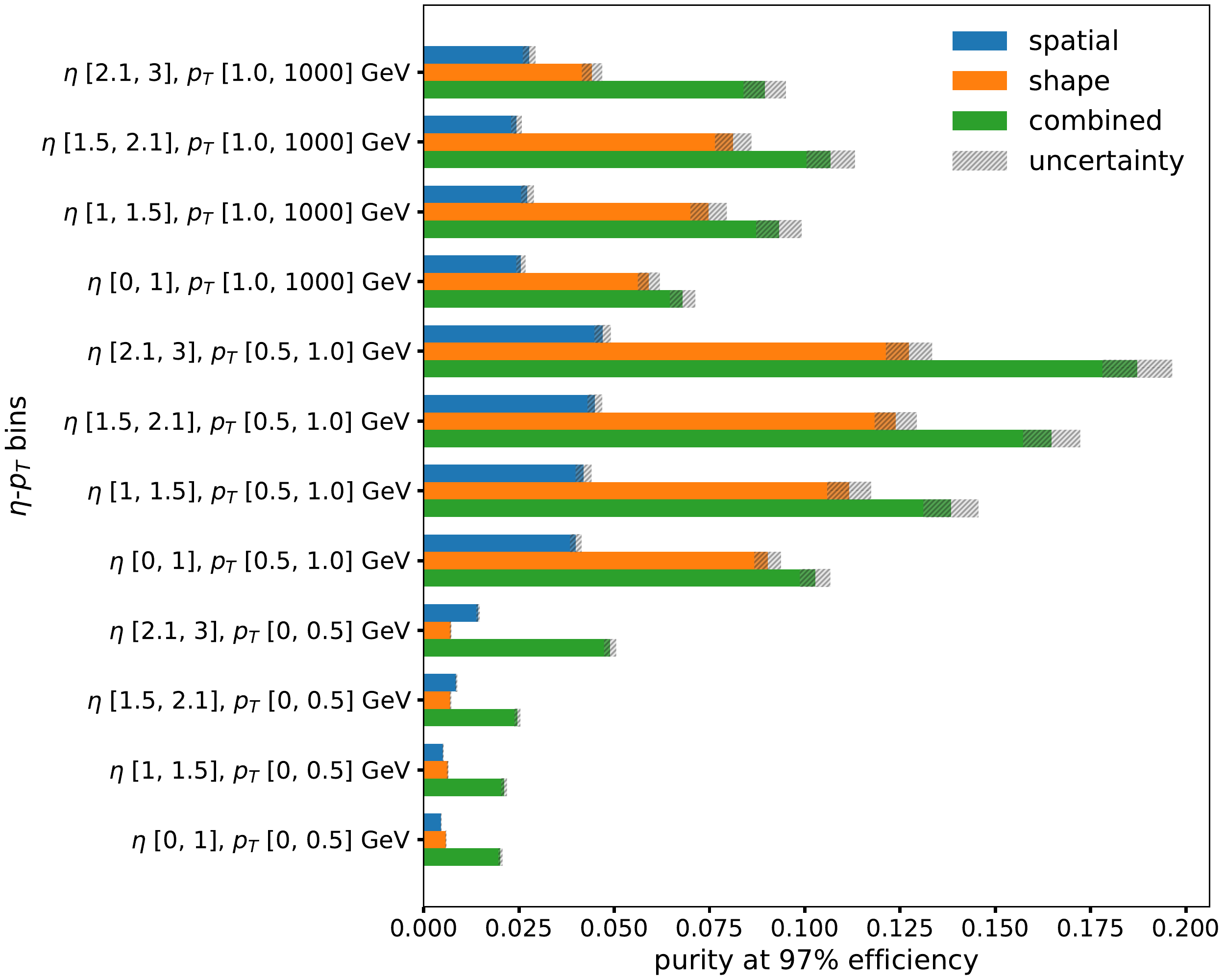}
    \caption{Comparison of the purity of the selected doublets for a given signal efficiency of 97\%
    in different $\eta$ and $\pt$ bins of the doublet candidates
    for the same neural network architecture trained with three different sets of input variables:
    only the hit locations in blue,  only the cluster shape information in orange and the combination of the two in green.  The gray band indicates the statistical uncertainty.}
    \label{fig:doublet_perf-purity}
\end{figure}

The neural network based selections are compared with the cut-based selections\footnote{$|\Delta\phi$/$\Delta r| < 0.0006$ rad/mm and $|z_0| < 100$ mm.} from Ref.~\cite{Ju:2020xty}, which have an efficiency and purity of $43\%$ and $3\%$, respectively.
This comparison is shown in Figure~\ref{fig:doublet_vs_cutbased}. For a purity of 3\%, using the neural network trained with cluster spatial information yields an efficiency that doubles the one by using the cut-based selection. Furthermore, adding the clustering information into the neural network increases the efficiency relatively by about 20\% to 0.98. The three differently trained neural networks are compared exclusively for doublets in different $\eta$ and \pt\ regions. The $\eta$ and \pt\ of each doublet are obtained from a helix fit of the doublet assuming the doublet starts from the origin. The helix fit is implemented with the conformal mapping method~\cite{Hansroul:1988wa}. Figure~\ref{fig:doublet_perf-purity} shows the comparison of the doublet purity for a given doublet efficiency of 97\% for the three doublet neural networks. For low-\pt\ doublets, the clustering information is as important as the spatial information in the central $\eta$ region and becomes more important in the high $\eta$ region. For all kinematic regions, combining the clustering information with the spatial information significantly boosts the doublet purity at this high 
doublet efficiency working point. 
Finally, the output from doublet network can be used to construct triplet clusters as described in the next section.

\subsection{Cluster Triplets}
\label{sec:threeclusters}
Triplets are composed of three hits with each coming from one of the three consecutive innermost layers\footnote{In practice, one could allow for gaps, but we consider the consecutive case for simplicity.}. As with the doublet case, the triplet neural network was trained to separate true triplets from fake triplets. 

As shown in the previous section, the seeding performance depends strongly on the seed kinematic properties.  Cluster triplets are sorted into $p_T$ and $\eta$ bins using the same helical fit, only now with three instead of two points.  For real triplets, the true momentum is known, but to directly compare with fake triplets, the same reconstruction algorithm is used in both cases.  Due to the large number of potential fake triplets, only a fraction of fake triplets from one event are used for training. Specifically, true triplets from about 1000 events are used for the signal and about 0.2\% of the possible fake triplets from one event are used for the background.  This results in approximately balanced classes for the training; for the final metrics, event weights are applied to reproduce the natural abundance of each class.

\begin{figure}[t]
    \centering
    \includegraphics[width=0.8\textwidth]{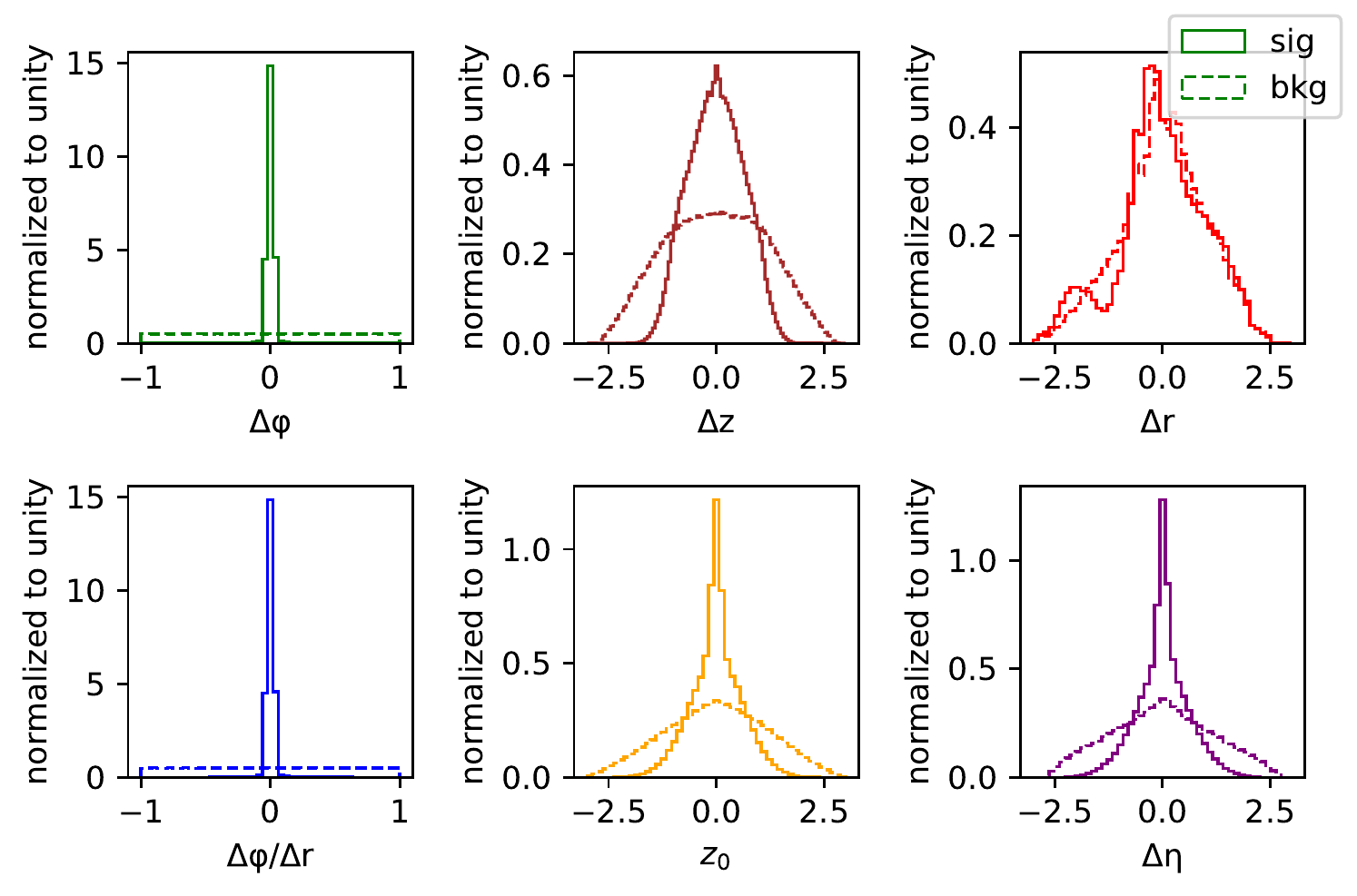}
    \caption{A subset of the spatial triplet features with all the parameters calculated from the first two consecutive layers. Solid line for true signal and dotted line for fake background.  Each feature is standardized to have zero mean and unit variance and thus are dimensionless.  
    \label{fig:spatial parameters distribution}}
\end{figure}

As with the doublets, three sets of features are used: `spatial', `shape', and `combined'. Each set of features are calculated from the same collection of three hits taken from three consecutive layers, with one hit coming from each layer. A spatial triplet is a group of twelve numbers: the first six entries are calculated by the differences in spatial coordinates of two hits on the first two consecutive layers whereas the second six entries are obtained in the same process but with two hits on the second two consecutive layers. Figure~\ref{fig:spatial parameters distribution} shows normalized histograms of each one of the first six entries of a spatial triplet.  The features are the same as the doublet case from the previous section.

The second set of features use only local information about the cluster shape. This information is summarized using the length of the clusters along the 
$v$ direction ($n_{ch1}$) in units of number of pixels.  Other features of the cluster shapes may be useful, but these features already contain significantly useful information.  The combination of the spatial and shape features gives 15 numbers for the `combined' feature set.  Each feature set is processed by a neural network of the same architecture.  These neural networks are composed of four layers with [128, 64, 32, 1] neurons and are trained with the binary cross entropy loss function. There are a large number of fake triplets that can be easily eliminated and make it difficult for the triplet network to learn.  In order to improve the training efficacy, we first apply a threshold on a doublet network.  In particular, we set a threshold on the doublet network so that we keep 99\% of the true doublets for each feature set. 

\begin{figure}
    \centering
    \includegraphics[width=0.8\textwidth]{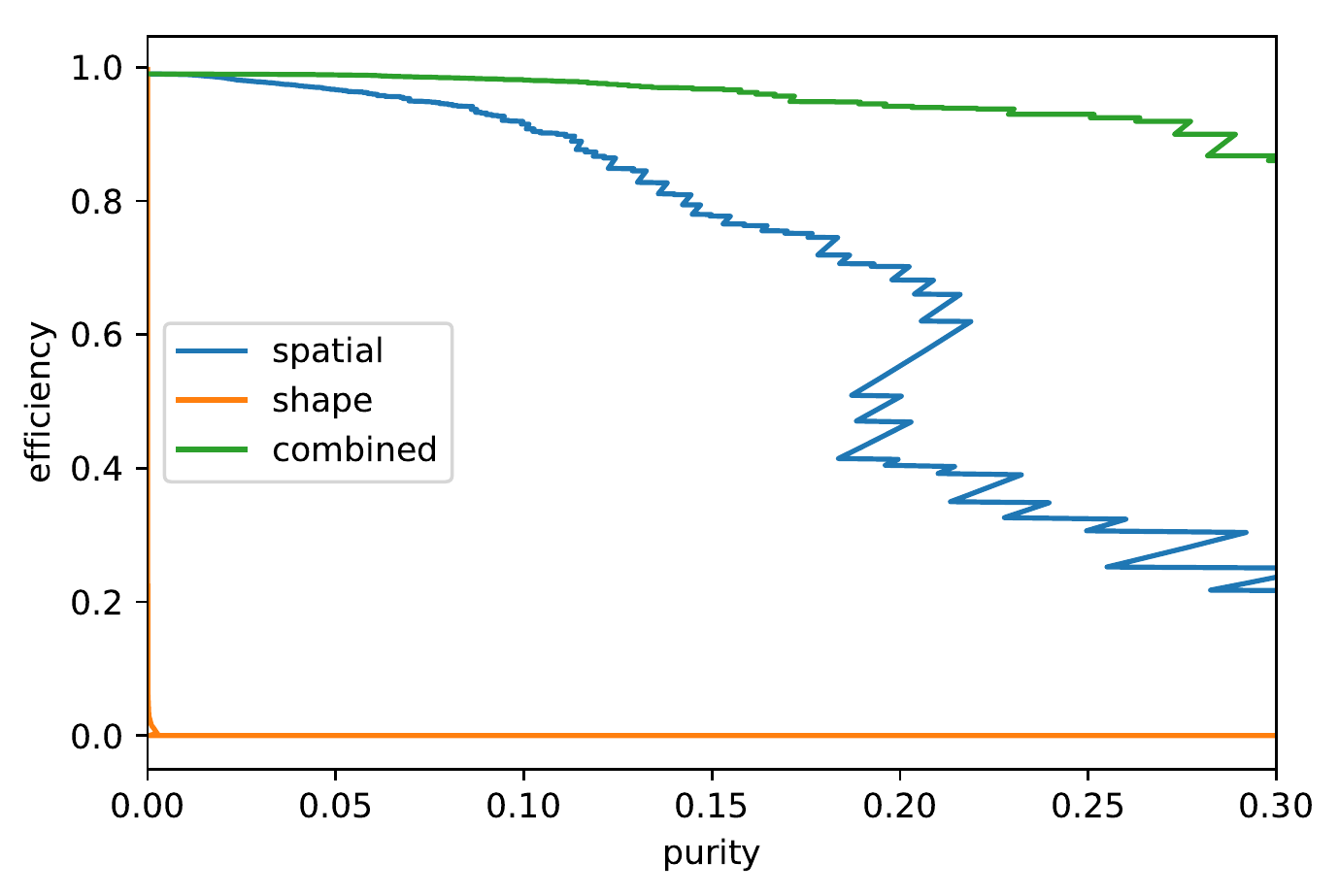}
    \caption{Efficiency versus purity for the three neural networks. Blue for the spatial features alone, orange for the shape features alone, and green for the combined features. At a fixed efficiency, the greater the purity the better the performance.
    \label{fig:purity vs efficiency}}
\end{figure}

Neural networks trained using the three different feature sets are compared in Fig.~\ref{fig:purity vs efficiency}.  For the evaluation of the purity and efficiency, weights are applied to correct the number of true and fake triplets to match their expected natural abundance.  The cluster information alone is not very effective at eliminating fake triplets, but the additional information significantly improves the efficiency for a fixed purity when combined with the cluster hit coordinates.  For example, at a purity of 30\%, the efficiency increases from about 20\% to nearly 90\%.  The jaggedness is a result of the limited statistics. 

\begin{figure}
    \centering
    \includegraphics[width=0.8\textwidth]{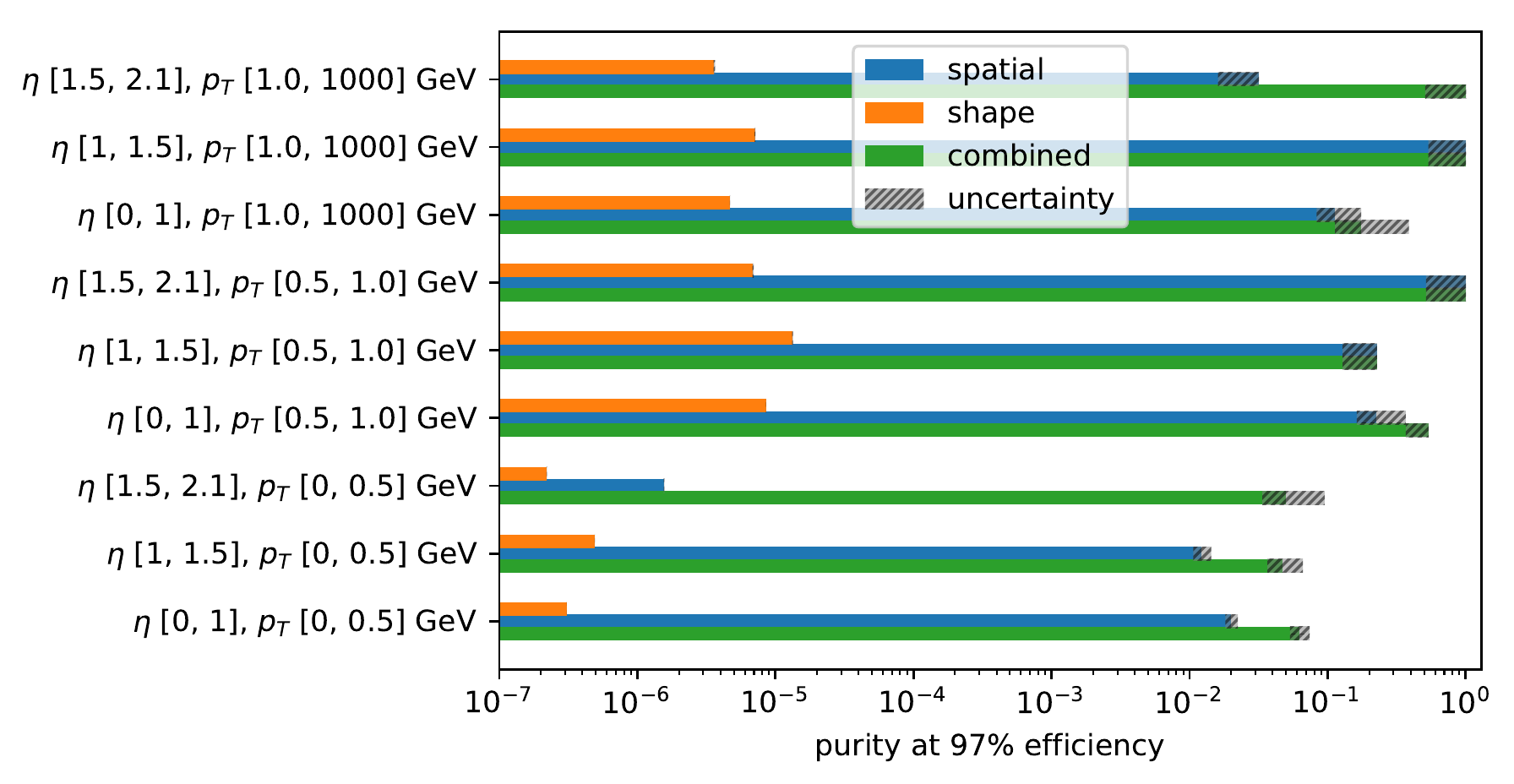}
    \caption{Comparison of the purity of the selected triplets for a given signal efficiency cut at 97\% in different $\eta$ and $\pt$ bins. Blue for the spatial feature alone, orange for the cluster feature alone, and green for the combined feature. The error bars represent the statistical uncertainty. High $\eta$ bins in the range of [2.1, 3] are not included due to the limited number of triplets passing the selection. }
    \label{fig:purity distribution}
\end{figure}

A differential perspective of the three networks is presented in Fig.~\ref{fig:purity distribution}, grouping the results in bins of seed $p_T$ and $\eta$.  For the fixed efficiency of 97\%, the combined feature set has the best purity except in three bins where the purity is statistically consistent with unity.  For each $p_T$ bin, all three triplet neural networks perform better at higher $|\eta|$ than at lower $|\eta|$ because the cluster lengths are longer and so the shape information is more useful.  Overall, the purity from networks trained only using cluster information is very low ($10^{-7}-10^{-5}$), but when combined with spatial information, the purity is everywhere above 1\% and in some cases is statistically consistent with 100\%. 
This improvement in the purity of triplet seeds should allow for a drastic improvement in overall track reconstruction time.      

\section{Conclusions and Outlook}
\label{sec:concl}

This paper has presented a series of studies that show how local pixel cluster shape information can be useful for performing track cluster seeding.  The additional information is useful when considering single clusters, pairs of clusters, as well as triplets of clusters.  For single pixel clusters, shape information is used to estimate the direction of the charged particle.  This vector estimate can be used to significantly reduce the number of hits in subsequent layers that need to be considered when building doublets and triplets.  As expected, this information is most useful when the pixel cluster is composed of more than one row in the $z$ direction.  Compared to using spatial information alone, the combined information reduces the number of hits inside a cone around the predicted trajectory by about a factor of ten.  For pairs of pixel clusters, one can use their relative orientation combined with the local cluster shapes to reject fake doublets that are not from a single charged particle.  The expected gains depend on momentum of the particles.  For example, for a signal efficiency of 97\%, the purity increases from about 0.5-1\% to 2-5\% at $p_T<500$ MeV and from about 3\% to 6-10\% for $p_T>1$ GeV.  There are also significant improvements for triplets of pixel clusters, with the most significant improvements projected to be at low $p_T$ and at high $|\eta|$ where many order of magnitude improvement in purity is possible at a fixed efficiency. Future studies could combine the approaches designed for single clusters, pairs of clusters, and triplets of clusters to further optimize the classification and computational performance.

There are a variety of extensions that would be interesting to pursue in the future.  For example, a direct comparison between the new approaches and traditional seeding algorithms would be an important step to understand the gains in practice.  The spatial baseline shown here uses all of the same information as traditional algorithms, but is expected to be better because it already uses a neural network to optimally process the hit positions.  With innovations in experiment-agnostic tools like \textsc{Acts}, a direct comparison may be possible in the near future and even larger gains than those shown here are expected.  Beyond comparing to traditional methods, it would also be interesting to extend the approaches here to include even more information (e.g. pixel cluster charge and precision timing).  It would also be interesting to apply these techniques to long-lived particles (LLPs).  The single cluster NN trained only on shape information is insensitive to the origin of the track.  On the other hand, all the other networks studied use information about the track origin and an extension to LLPs would require more study, with an appropriate data set. 
Innovations in deploying neural networks on accelerators like FPGAs~\cite{Duarte:2018ite} could also make the implementation of these seeding algorithms fast enough to fit into trigger time budgets.  
These innovations and more may significantly improve track seeding for the HL-LHC era and beyond.

\section*{Acknowledgments}

We would like to thank Maurice Garcia-Sciveres for many useful discussions and Claudius Krause, Jack Collins, and Zhen Liu for collaboration in the early stages of this project.  We also thank Maurice, Claudius, Jack, and Zhen for detailed feedback on the manuscript.  This work was supported by the U.S.~Department of Energy, Office of Science under contract DE-AC02-05CH11231. This manuscript has been authored by Fermi Research Alliance, LLC under Contract No. DE-AC02-07CH11359 with the U.S. Department of Energy, Office of Science, Office of High Energy Physics.

\bibliographystyle{jhep}
\bibliography{myrefs.bib}{}

\end{document}